# First Principle QCD QED Potentials, Quark Confinement and Electron-Positron Pair Annihilation

Eue-Jin Jeong
The University of Texas at Austin
Austin TX 78712

## Abstract

The mystery of the origin of the mass of the elementary particles persists despite the success of quantum field theories to the highest level of accuracy. Renormalization program has been the essential part of the calculation of the scattering amplitudes where the infinities of the calculated mass of the elementary particles are subtracted for progressive calculation of the higher order terms. The mathematical structure of the mass from quantum field theories expressed in the form of infinities suggests that there exists finite dynamical mass in the limit the input mass parameter becomes zero. The Lagrangian recovers symmetry at $m \to 0$ while the self energy diagrams acquire finite dynamical mass. The first principle QCD and QED potentials are obtained by replacing the fixed mass and coupling constant in Yukawa potential with the scale dependent running coupling constant and the corresponding mass which predicts quark confinement and electron-positron pair annihilation respectively.

## Dynamical Mass from the Massless Quantum Field Theory

The standard Glashow-Weinberg-Salem (1) model of the electroweak interaction has been highly successful in predicting the interactions of high energy elementary particles. The discovery (2) of the W and Z gauge bosons, and finally the discovery of Higgs boson at CERN in 2012 (3) proved that Standard Model is a mathematically correct theory describing the interactions of the elementary particles.

We investigate the structures of the self energy diagrams of the elementary particles to study the relationship between the mass and coupling constant in quantum field theories. By using the dimensional regularization method for the renormalization of the quantum field theories, finite indeterminate mathematical form of the dynamical mass of the fields is obtained in the limit the input mass term in the Lagrangian approaches to zero. In this process, the symmetry of the original Lagrangian is restored while the finite mass appears in the self energy loop diagrams.

The renormalization group equation (4) resolved the problem of the arbitrariness of the renormalization prescription. In the following, the dynamical mass generation mechanism is presented within the framework of the dimensional regularization method developed by G. 't Hooft and M. Veltman (5).

**Dynamical Mass**

(1) $\lambda\phi^4$ theory

The mathematical structure of the one loop self energy diagram in $\lambda\phi^4$ theory is represented by

$$\text{One loop} = \frac{m_0^2 \lambda}{16\pi^2}\left[\frac{1}{n-4} + \frac{1}{2}\psi(2) - \frac{1}{2}\ln\frac{m_o^2}{4\pi\mu^2} + O(n-4)\right], \qquad [1]$$

where $\psi(2)$ is a constant given in general,

$$\psi(n+1) = 1 + \frac{1}{2} + \ldots + \frac{1}{n} - \gamma \quad (\gamma = 0.5772\ldots)$$

and $\mu$ is an arbitrary constant with the dimension of mass. The renormalization for the nonzero bare mass $m_0$ is necessary because the first term is divergent in the $n \to 4$ limit. However, in the zero bare mass limits $m_0 \to 0$ the term is not infinity but becomes undetermined. We introduce a constant $C_s$, and the one loop diagram becomes

$$\lim_{\substack{m_0 \to 0 \\ n \to 4}}\left[m_0^2 \frac{\lambda}{16\pi^2}\left[\frac{1}{n-4} + \frac{1}{2}\psi(2) - \frac{1}{2}\ln\frac{m_o^2}{4\pi\mu^2} + O(n-4)\right]\right] = \lim_{\substack{m_0 \to 0 \\ n \to 4}}\frac{\lambda}{16\pi^2}\frac{m_0^2}{n-4} \qquad [2]$$

$$\equiv \frac{\lambda}{16\pi^2}C_s,$$

where $C_s = \lim\limits_{\substack{m_0 \to 0 \\ n \to 4}} \frac{m_0^2}{n-4}$.

As a result of this operation, we have analytical mass which is not infinity. The massless $\lambda\phi^4$ scalar field theory starts to have mass from one loop self energy diagram. Recalling that the $\lambda\phi^4$ massless scalar field theory is the simplest case of the supersymmetric theories, it provides us with a clue to the possible mass generation mechanism for the supersymmetric particles.

The fact that the explicit mass parameter in the Lagrangian does not represent the real mass of the field and its sole purpose is to provide a reference from which the real mass is determined by experiment after renormalization has already suggested that mass can be generated by dynamical interactions by the interacting fields. In the case of QCD and QED, self energy is calculated without the explicit mass parameters in the Lagrangian.

(2) QED

The self energy diagram of the electron in QED (7) without the mass parameter in the Lagrangian is given by

$$\Sigma(p) = \frac{2}{n-4}\frac{e^2}{16\pi^2}P - \frac{e^2}{8\pi^2}\left[\frac{1}{2}P(1+\gamma) + \int_0^1 dx P(1-x)\ln\frac{p^2 x(1-x)}{4\pi\mu^2}\right] \quad [3]$$

where P represents the energy-momentum tensor of the electron. Since the self energy is defined by the energy when the particle is at rest state, the mass of the electron is given by,

$$M_e = \frac{e^2}{8\pi^2}C_e \quad [4]$$

where $C_e = \lim_{\substack{p \to 0 \\ n \to 4}} \left[\frac{\det P}{n-4}\right]$.

The vacuum polarization diagram of the photon (7) is given by

$$\Pi_{\mu\nu}(p) = \frac{e^2}{2\pi^2}\left(P_\mu P_\nu - \delta_{\mu\nu}P^2\right)\left[\frac{1}{3(n-4)} - \frac{1}{6}\gamma - \int_0^1 dx x(1-x)\ln\frac{p^2 x(1-x)}{2\pi\mu^2}\right] + O(n-4) \quad [5]$$

The dynamical mass of the photon is now given by $M_\gamma^2 = \frac{e^2}{6\pi^2}C_\gamma$

, where the photon mass constant is $C_\gamma = \lim_{\substack{p \to 0 \\ n \to 4}}\left[\frac{\det(P_\mu P_\nu - \delta_{\mu\nu}P^2)}{n-4}\right]$. While it is generally known that photons do not carry mass, the gauge invariance of the Lagrangian suggests that they manifest mass in relation to the distance of their interactions with the charged particles.

## Relationship between the Mass and the Coupling Constant

It is well known that the electron's mass is related to the electrostatic self energy in classical electrodynamics, in which the radius $r_0$ of the electron is defined by

$$m_e = \frac{e^2}{r_0}. \qquad [6]$$

In fact, the relationship between the mass and the corresponding charge of a particle is a universal feature beyond the classical electrodynamics. The quantum field theoretical mass of the quantum fields are directly related to the corresponding coupling constants by the following relations as shown from the above examples.

$$M_s^2 = \frac{\lambda}{16\pi^2} C_s \qquad C_s = \lim_{\substack{m_o \to 0 \\ n \to 4}} \left[ \frac{m_o^2}{n-4} \right] \qquad [7]$$

$$M_e = \frac{e^2}{8\pi^2} C_e \qquad C_e = \lim_{\substack{p \to 0 \\ n \to 4}} \left[ \frac{\det P}{n-4} \right] \qquad [8]$$

$$M_\gamma^2 = \frac{e^2}{6\pi^2} C_\gamma \qquad C_\gamma = \lim_{\substack{p \to 0 \\ n \to 4}} \left[ \frac{\det(P_\mu P_\nu - \delta_{\mu\nu} P^2)}{n-4} \right] \qquad [9]$$

$$M_f = C_3 \frac{g^2}{8\pi^2} C_f \qquad C_f = \lim_{\substack{p \to 0 \\ n \to 4}} \left[ \frac{\det P}{n-4} \right] \qquad [10]$$

$$M_{Y.M}^2 = \left( \frac{5}{3} C_1 - \frac{4}{3} C_2 \right) \frac{g^2}{8\pi^2} C_{Y.M} \qquad C_{Y.M} = \lim_{\substack{p \to 0 \\ n \to 4}} \left[ \frac{\det(P_\mu P_\nu - \delta_{\mu\nu} P^2)}{n-4} \right] \qquad [11]$$

where $C_1$, $C_2$, $C_3$ are constants determined by the group structure of the nonabelian gauge theory and the sub-indices s, e, $\gamma$, f, Y.M, indicate scalar, electron, photon, fermion and Yang-Mills field respectively. In the four dimensional space, all the constants for the self energies become undetermined in the limit the momentum and the input mass becomes zero. These relations between the mass and the coupling constant suggest a significant variation of the mass due to the running coupling constant.

## Generalized Yukawa Potentials for QCD QED and Quark Confinement

In 1935, H. Yukawa (6) introduced the nuclear potential which has been proven highly successful in addressing many of the diverse nuclear interactions. The major property of Yukawa's potential is the introduction of the mass of the pion as the interaction mediating particle which works for strong nuclear force at short distances. The coupling constant and the mass of the pion in Yukawa's nuclear potential are independent fixed parameters from each other

$$V_{Yukawa}(r) = -g^2 \frac{e^{-\alpha mr}}{r} \qquad [12]$$

where *g* is a magnitude of the coupling constant and *m* is the mass of the intermediate particle and *r* is the radial distance to the particle, and *α* is a scaling constant.

Since we have established the dependence between the scale dependent coupling constant and the mass, we suggest constructing the new generalized Yukawa potential by replacing the fixed mass and coupling constant with the ones that depend on the running coupling constant. Using the mathematical relationship between the mass and the quantum field theoretical interactions [7]-[11], the fixed mass and the coupling constant in Yukawa potential are replaced by the scale dependent running coupling constant and the corresponding mass as following.

$$V(r) = g^2(\mu) \frac{e^{-\alpha m(\mu)r}}{r} \qquad [13]$$

where $g^2(\mu)$ is the running coupling constant and $m(\mu)$ is the scale dependent mass of the interaction mediating field in QFT. Using the scale dependent running coupling constant from QCD

$$g^2(\mu) = \frac{g_0^2}{1 + \frac{g_0^2}{8\pi^2} \ln \frac{\mu}{\mu_0}} \qquad [14]$$

which was developed by D. Gross, F. Wilzeck and H. D. Politzer (8) (9) and the scale dependent rest mass of Yang-Mill field [11], the QCD potential is given by

$$V_{qcd}(r) = \frac{g_0^2}{1 + \frac{g_0^2}{8\pi^2} \ln \frac{\mu}{\mu_0}} \frac{1}{r} \exp\left(-\left(\frac{\frac{\alpha}{8\pi^2} g_0^2 C_g C_k}{1 + \frac{g_0^2}{8\pi^2} \ln \frac{\mu}{\mu_0}}\right)^{1/2} r\right) \qquad [15]$$

The potential in the form [15] is not practical due to the parameter $\mu$ which depends on the input momentum. To translate $\mu$ into the distance *r*, we suggest that there is a mathematical relationship between $\mu$ and *r* governed by

$$\mu = \exp\left(\frac{\rho}{r^2}\right) \quad \rho > 0 \qquad [16]$$

While the relation [16] does not violate the quantum mechanical uncertainty, since the larger input momentum $\mu$ results into a smaller distance $r$ due to the quantum uncertainty principle,

$$\Delta x \Delta p \geq \hbar/2 \qquad [17]$$

justification for the relation depends on the predictability of the subsequent results. In fact, it turns out that the mathematical relation [16] is the only possible choice to bring out both QCD and QED interaction potentials in confirmation of the phenomenological quarkonia spectroscopy results (13).

In the limit $\frac{g_0^2}{8\pi^2}\ln\frac{\mu}{\mu_0} \gg 1$, the QCD potential [15] is given by

$$V_{qcd}(r) = \frac{1}{\frac{A}{r^2}-B}\frac{1}{r}\exp(-(\frac{\frac{\alpha}{8\pi^2}C_k C_g}{\frac{A}{r^2}-B})^{1/2} r) \qquad [18]$$

, where $A = \rho\frac{g_0^2}{8\pi^2}$, $B = \frac{g_0^2}{8\pi^2}\ln\mu_0$ and $C_g$ is the gluon mass constant which is given by $C_g = 5.8\times 10^{-103} g^2$ using the upper limit of the photon rest mass $3\times 10^{-53} g$ (10) and $C_\kappa$ is a group structure constant having the order of magnitude 1.

For $\alpha=1$, the potential has a peak at $r \approx \sqrt{\frac{A}{B}}(1-3A^2 C_g)$ and its magnitude is given by

$$V_{peak} = \frac{8\pi^2}{6C_g B^{1/2} A^{5/2}}\exp(-\frac{1}{B\sqrt{6A}}) \qquad . \qquad [19]$$

At $r = r_{cq} = \sqrt{\frac{A}{B}}$, the potential [18] is damped to zero by the negative infinity exponential term and becomes imaginary as $r$ becomes larger.

In quantum mechanics, imaginary potential is known to violate the conservation of the probability of finding the quantum particles. The loss of probability beyond the outer radius of the hadronic boundary is consistent with the prediction of Hawking radiation (11) assuming that the blackhole is fundamentally a neutron star with extreme density of quark-gluon plasma.

For small $r$ and $\alpha=1$, the quark potential [18] becomes

$$V_{qcd}(r) = \frac{r}{A}(1-\frac{B}{A}r^2 + ......) \qquad [20]$$

The quark potential [18] has the following features.

1. linear potential at small distance $r$
2. confinement at ordinary energy level
3. becomes imaginary beyond the critical distance of the hadronic boundary at high energies
4. no singularity throughout the relative distances

By applying the same mathematical procedure for the case of QED using the running coupling constant for QED

$$\gamma(Q) = \frac{\gamma}{1 - \frac{2\gamma}{3\pi} \ln \frac{Q}{m_e}} \qquad [21]$$

where $\gamma = \frac{e^2}{4\pi}$ and $Q = \exp(\frac{\rho}{r^2})$ [16], the QED potential is given by

$$V_{qed}(r) = \frac{-1}{C - \frac{D}{r^2}} \frac{1}{r} \exp(-(\frac{\frac{\alpha}{6\pi^2} C_\gamma}{C - \frac{D}{r^2}})^{1/2} r) \qquad [22]$$

where $C = 1 + \frac{e^2}{6\pi^3} \ln m_e$, $D = \frac{e^2 \rho}{6\pi^3}$ and $C_\gamma$ is the photon mass constant [9].

The QED potential [22] reduces into the usual Coulomb potential in the intermediary distances and as $r$ becomes larger the potential approaches to zero fast due to the negative exponential factor. As the distance decreases toward $\sqrt{\frac{D}{C}}$, the potential has a peak depth but not infinity. As $r$ becomes smaller than $r_{ce} = \sqrt{\frac{D}{C}}$, the potential shows an abrupt core rising up to zero level and becomes imaginary. This behavior of the QED potential is in accordance with the electron-positron pair annihilation as they approach sufficiently close together. Also, the sharply rising core potential has been employed for the calculation of the "Lamb shift" (12) in the form of the $\delta(r)$ function.

By adding the two potentials [18] and [22] at the intermediary distance, we have

$$V(r) = \frac{-1}{Cr} + \frac{r}{A} \qquad [23]$$

This result confirms the previously reported non-relativistic quark potential which was in good agreement with the experimental results in the heavy quarkonia spectroscopy (13).

The quark potential [18] shows a small yet finite probability of finding fractional charges beyond the critical distance which supports the results reported by many researchers (14), even though the quark itself has not been isolated. In the case of the QED potential with electron-positron annihilation, the interaction of the electron with the antimatter positron is considered the key to the loss of the electron's quantum probability at close distances. Yet, it is a mystery that the interaction potential derived from the exchange of space-like photons predicts electron-positron annihilation by displaying the imaginary potential at close distances. On the other hand, in the case of QCD, the quark's loss of quantum probability beyond the distance of the hadronic boundary despite the apparent absence of nearby anti-quark matter is considered a fundamental mystery although it confirms the experimental data and is consistent with Hawking radiation [11] in the case of the blackhole.

We presented the uniform mathematical procedure to transform the perturbative quantum field theories into unified interaction potential model for both QCD and QED by utilizing the running coupling constant derived from the renormalization group equations within the framework of the known Yukawa nuclear potential model.

## Acknowledgement

Quantum Field Theories are riddled with the problems of dealing with the infinities that have to be systematically subtracted to obtain successful theoretical predictions which have been a cause of skepticism in the development of the modern theory of the elementary particle physics. We resolved the issue by establishing the mathematical connection between the QFT predictions of the running coupling constant and the new theoretical potential models for the elementary particle interactions by quarks/ gluons and electrons/photons. It was a great pleasure to learn the dimensional regularization method developed by M. Veltman and G. 't Hooft for the renormalization of the quantum field theories while Dr Veltman's visit to the University of Michigan Ann Arbor at the time.